\documentclass[a4paper,11pt]{article}
\pdfoutput=1 % if your are submitting a pdflatex (i.e. if you have
\usepackage{jinstpub} % for details on the use of the package, please
\usepackage{amsmath}

\usepackage{siunitx}

\usepackage{graphicx}

\usepackage{subcaption}

\usepackage{lineno}
\title{\boldmath Comparison of Silvaco and Synopsys
TCAD Predictions Including the
Perugia Radiation Damage Model in
Silicon Pixel Detectors for the HL-LHC}

\author[a,1]{M. Bomben,\note{Corresponding author.}}
\author[b,c]{T. Croci,}
\author[b]{K. Aouadj,}
\author[b,d]{A. Fondacci,}
\author[b,e]{F. Moscatelli,}
\author[b]{A. Morozzi,}
\author[b,d]{D.~Passeri}
\affiliation[a]{APC, Universit\'e Paris Cite, CNRS/IN2P3, 10 rue Alice Domon et Leonie Duquet, Paris, 75013, France}
\affiliation[b]{Istituto Nazionale di Fisica Nucleare, Sezione di Perugia, via A. Pascoli, 06123, Perugia, Italy}
\affiliation[c]{Dipartimento di Fisica e Astronomia, Universit\`a degli Studi di Padova, via F. Marzolo 8, 35131, Padova, Italy}
\affiliation[d]{Engineering Department, University of Perugia, via G. Duranti 93, 06125 Perugia}
\affiliation[e]{Consiglio Nazionale delle Ricerche, Istituto Officina dei Materiali, via A. Pascoli, 06123, Perugia}
\emailAdd{marco.bomben@cern.ch}

\abstract{At the High Luminosity Large Hadron Collider (HL-LHC), silicon pixel detectors will be exposed to radiation fluences about 5 to 10 times larger than those experienced by the current innermost pixel 
layers up to today. Due to radiation damage to bulk of pixel detectors, leakage current and depletion voltage will increase significantly over time, posing 
severe constraints on operating conditions, with important modifications to the electric field profile due to radiation induced deep defects in silicon bulk. It is important to have reliable predictions for all observables - such as leakage current level and breakdown voltage - after irradiation, in 
order to estimate operational voltage values. In this paper, the predictions of Silvaco and Synopsys TCAD device simulations are compared when the surface and 
bulk defects and traps of the ``New University of Perugia radiation damage model'' are included by studying observables like leakage current, depletion and breakdown voltage, along with electric field and trap occupancy profiles.
The results are quite promising regarding leakage current, depletion voltage, electric field and trap statistics, at two distinct reference temperatures and fluences.}

\keywords{Detector modelling and simulations, Simulation methods and programs}

\arxivnumber{2604.26386} % only if you have one

\proceeding{21$^{\text{st}}$ Workshop on Advanced Silicon Radiation Detectors\\
  17-19 February 2026\\
  Perugia, Italy}

\begin{document}
\maketitle
\flushbottom

\section{Introduction}
\label{sec:intro}

In the next few years the CERN Large Hadron Collider (LHC)~\cite{LHCTDR} will become a high luminosity machine, the High-Luminosity LHC (HL-LHC)~\cite{HLLHCTDR}, with instantaneous 
luminosity increasing by a factor 2 to 4 and the goal of accumulating 10 times more data in about 10 years of operation. Together with luminosities, hit rate and radiation damage will increase as well, 
with the innermost part of the tracking detectors seeing a 5 to 10 fold increase in fluence. Both ATLAS~\cite{ATLAS} and CMS~\cite{CMSDetector} experiments will undergo upgrades to cope with the much 
harsher environment of the High Luminosity phase. In particular the tracking detectors~\cite{AtlasID1,AtlasID2,Karimaki:368412} of the two experiments will be 
upgraded~\cite{ATLASITkPixelTDR,ITkStripsTDR,CMSTrackerUpgradeTDR}. Both systems will be instrumented with hybrid silicon detectors, pixel ones in the core part and microstrips at larger radii.

The high energetic hadrons produced in the $p-p$ collisions of HL-LHC will damage the silicon material both in the bulk and at the interface to dielectrics. As a result, deep defects~\cite{MollTNS2018} 
are created which are responsible of macroscopic effects like the increase of leakage current, the change of the operational voltage, signal loss and modifications of the behaviour of front-end 
electronics ~\cite{Dawson:2764325}. Signal loss and the related modification of pixels cluster shapes are the most impacting effect to higher level objects like tracks. It is important to have precise 
simulations of sensors response in order to anticipate potential performance losses and to train reconstruction tools on simulated events samples that are radiation damage 
aware~\cite{Swartz:2008oU,EPS2023Bomben,BattagliaEPS2023} as it is done for the actual tracking systems.

TCAD\footnote{Technology Computer Aided Design} simulation tools allow to model the electric field modifications inside the silicon bulk and at the interface with dielectrics by including effective 
defects making it possible with limited simulation burden to reproduce evolution of the sensors response with the accumulation of radiation dose and fluence. In this paper a comparison of 
simulation results from two different tools - Synopsys~\footnote{https://www.synopsys.com/manufacturing/tcad.html - device simulation software version: V-2023.12} and Silvaco~\footnote{https://silvaco.com/tcad/ - device simulation software version: 1.24.0.R} TCAD - is reported for 
an n-on-p diode under radiation damage conditions using a limited number of effective defects. Demonstrating agreement between the results of these two tools will 
provide mutual cross-validation and further substantiate the soundness of the implemented radiation damage modelling.
The radiation damage induced effects are simulated by the implementation of the University of Perugia bulk~\cite{NewPGBulk2024} and surface~\cite{PGSurf2019} damage model 
(``New University of Perugia'' model).  
After introducing them in section~\ref{sec:models}, the simulation setup (section~\ref{sec:setup}) is discussed; the simulated scenarios are observables are presented in section~\ref{sec:scenarios}. 
Simulation studies have been performed before (\ref{sec:unirr}) and after irradiation (\ref{sec:irr}). Section~\ref{sec:concl} have the conclusions.

\section{Radiation Damage Models}
\label{sec:models}

The University of Perugia group developed radiation damage models for TCAD device simulation tools for Silicon radiation sensors using Synopsys Sentaurus TCAD tool. The model consists of two parts, 
one for the radiation damage in the sensor bulk and the other for the damage to surface. The bulk model~\cite{NewPGBulk2024}, intended for high resistivity p-type silicon, 
features two acceptor defects and a donor one with densities proportional to the radiation damage fluence $\Phi$; in addition, to model the acceptor creation in the bulk, the bulk concentration $N_A$ (in cm$^{-3}$) 
is modelled by a fluence dependent function reported in equation~\ref{eq:bulkmod}:
\begin{equation}
\label{eq:bulkmod}
 N_{A} =
    \begin{cases}
      N_{A}(0) +g_c\Phi & 0 \le \Phi \le3\times10^{15} \, \SI{}{n_{eq}/cm^{-2}}\\
      4.17\times 10^{13}\cdot \ln{\Phi}-1.41 \times 10^{15}  & \Phi > 3\times10^{15}\, \SI{}{n_{eq}/cm^{-2}}
    \end{cases}   
\end{equation}
where $N_A(0)$ is the initial acceptor density and $g_c$ = 0.0237~cm$^{-1}$.  The surface damage model~\cite{PGSurf2019}  features a continuous distribution of defects in energy, acceptor-like in the upper 
half of the energy bandgap and donor-like in a range starting from the upper edge of the valence band covering 0.60~eV in energy; the surface model includes also the linear increase of oxide charge with 
dose.

\section{Simulation Setup}
\label{sec:setup}

\begin{figure}[htbp]
   \centering
   \includegraphics[height=0.335\textheight]{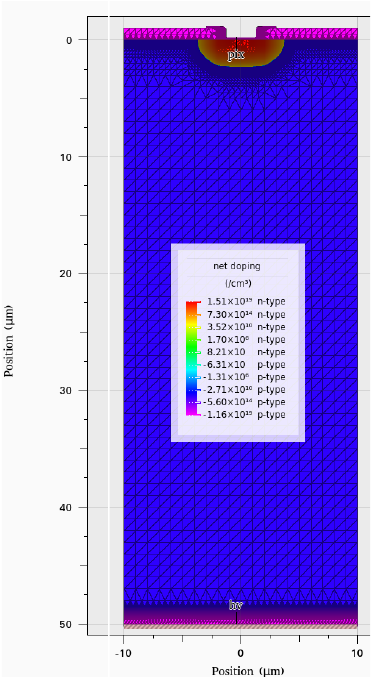} 
   \includegraphics[width=0.49\textwidth]{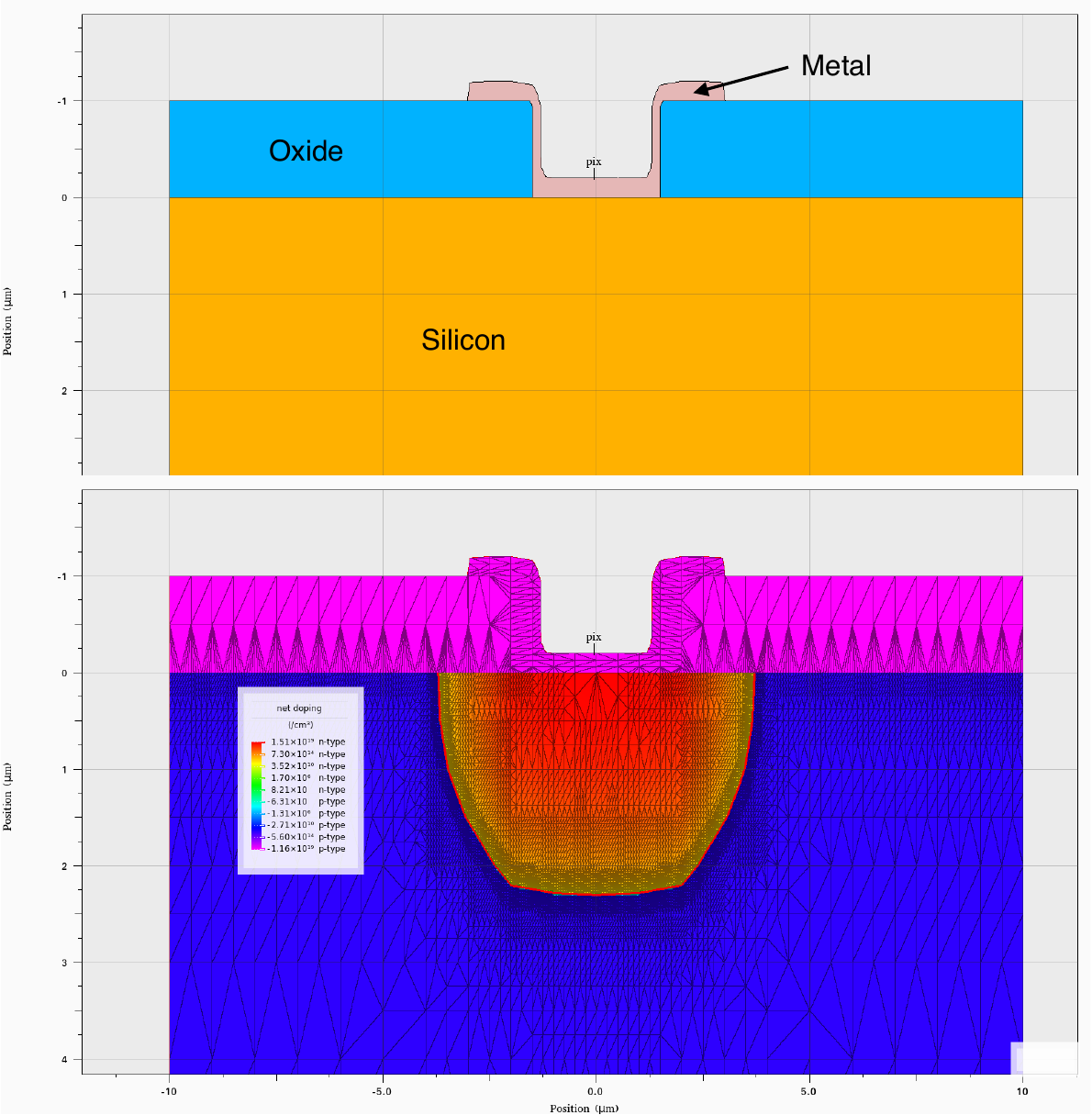} 
      \caption{\label{fig:structure}Simulated structure. (left) Whole structure, showing the net doping concentration and the mesh grid. (right) Zoom on the junction region: (top) the structure close to the 
      collection electrode (``pix'') ; (bottom) the net doping concentration and the mesh grid for the same region, with the red line indicating the junction position. }
   \end{figure}

A n-on-p diode was used to test the porting of the University of Perugia radiation damage model from Synopsys to Silvaco TCAD. The diode was \SI{50}{\micro m} thick with a Boron bulk concentration of 
$5 \times 10^{12}$~cm$^{-3}$. The structure was simulated in 2D with a lateral extent of \SI{20}{\micro m} and a small collection electrode; the junction side of the sensor was covered with a \SI{1}{\micro m} thick oxide 
layer. 
In Figure~\ref{fig:structure} details of the simulated structure are reported.

The device simulation setup for both TCAD tools included the following physics models: 
\begin{itemize}

\item statistics: Fermi-Dirac
\item recombination: Scharfetter - Shockley Read Hall~\cite{Scharfetter}
\item band-to-band tunneling: Schenk~\cite{SCHENK19921585}
\item bandgap narrowing: Slotboom~\cite{SLOTBOOM1976857}
%\item mobility model: Schwarz and Russe~\cite{1483325}
\item impact ionisation: van Overstraeten and de Man~\cite{VANOVERSTRAETEN1970583} with the gradient of the quasi-Fermi level as driving force\footnote{By default, the driving force is set to
the electric field in Silvaco and to the gradient of the quasi-Fermi level in
Synopsys.}
\end{itemize}

%Concerning the impaction ionisation model it has to be noted that in the original work of the University of Perugia group the Massey model~\cite{Massey} was used; the latter at the time of the 
%writing of this manuscript is still not available in Silvaco so it was decided to use the van Overstraeten and de Man one.

\section{Simulated Scenarios and Observables}
\label{sec:scenarios}

The study involved simulating the device before and after irradiation - $\Phi = 1.5\times10^{15}$~\SI{}{n_{eq}cm^{-2}} and $1.0\times10^{16}$~\SI{}{n_{eq}cm^{-2}}; two temperature values have been 
considered, $T = $~248~K and 300~K. Observables from simulation results included: current-voltage (IV) and capacitance-voltage (CV)\footnote{a frequency $f$=~1~kHz was used} characteristics, electric field and trap ionisation probabilities profiles.

\section{Unirradiated Device: Results}
\label{sec:unirr}

\begin{figure}[htbp]
   \centering
   \includegraphics[width=0.49\textwidth]{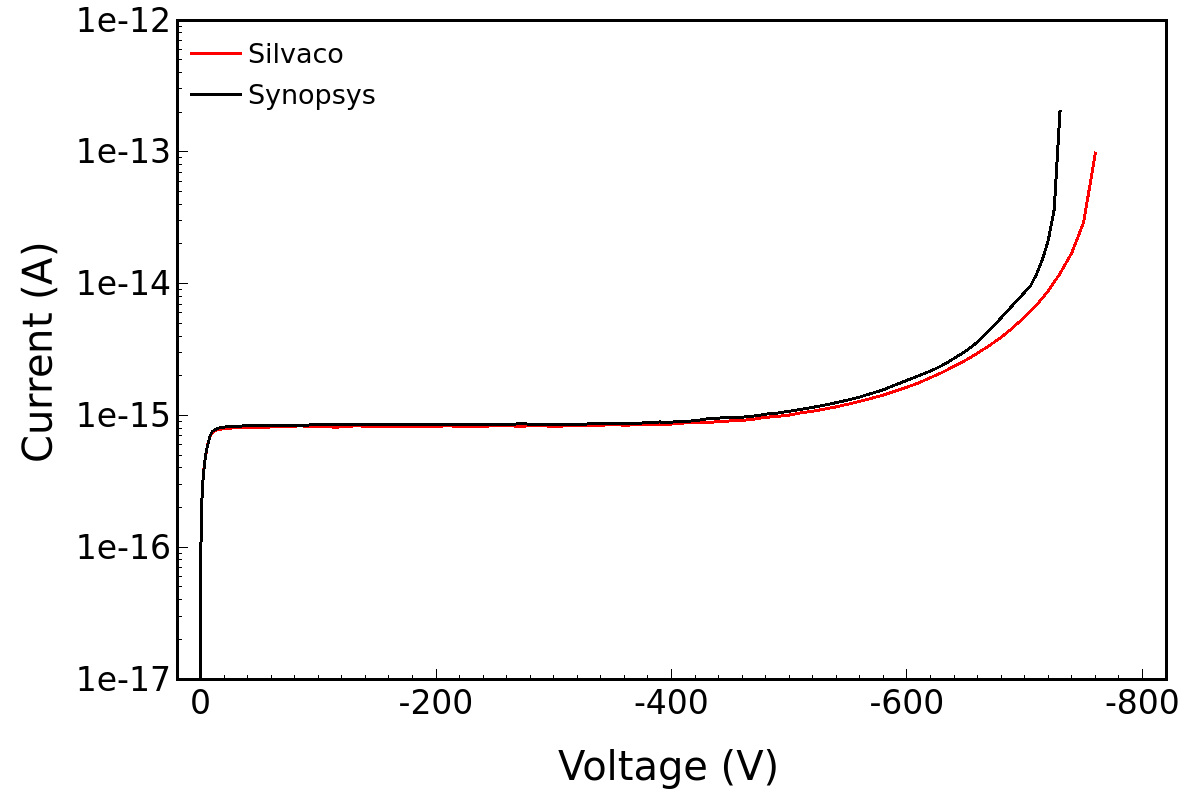}
   \includegraphics[width=0.49\textwidth]{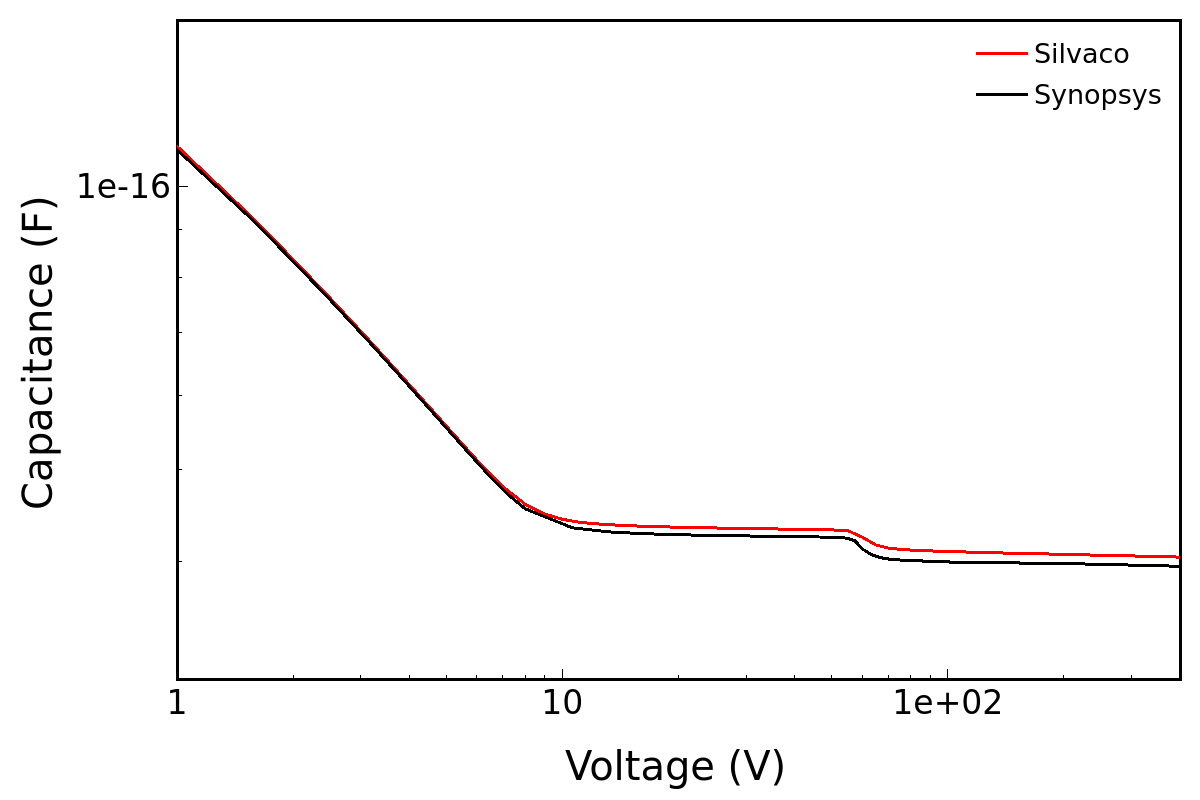} % requires the graphicx package
   \caption{\label{fig:unirr}Results from unirradidated device. (left) IV curve, (right) log(C) vs log (V) curve.}
\end{figure}

In Figure~\ref{fig:unirr} the main results from the simulation of the unirradiated device are presented. The diode was simulated at $T = $~300~K, including only interface charges and excluding interface traps in 
the surface damage model.

For what concerns the IV characteristics, the agreement between the tools is practically perfect  up to about 500 V; for larger voltages Synopsys predicts larger current than Silvaco, indicating an earlier 
breakdown voltage - around 700~V - while Silvaco predictions is at about 730~V. The origin of this discrepancy could lie in the different mesh between the two tools. 

For what concerns the CV characteristics, the two tools correctly predict a depletion voltage around 10~V and they overall agree allover the voltage range investigated, including the ``step'' observed at around 
55~V, which is probably due to the depletion below the oxide not covered by the collection electrode.

\section{Irradiated Device: Results}
\label{sec:irr}

In the following the results from the simulations of the irradiated device will be presented when both bulk and interface defects are simulated. 

\begin{figure}[htbp]
   \centering
   \includegraphics[width=0.49\textwidth]{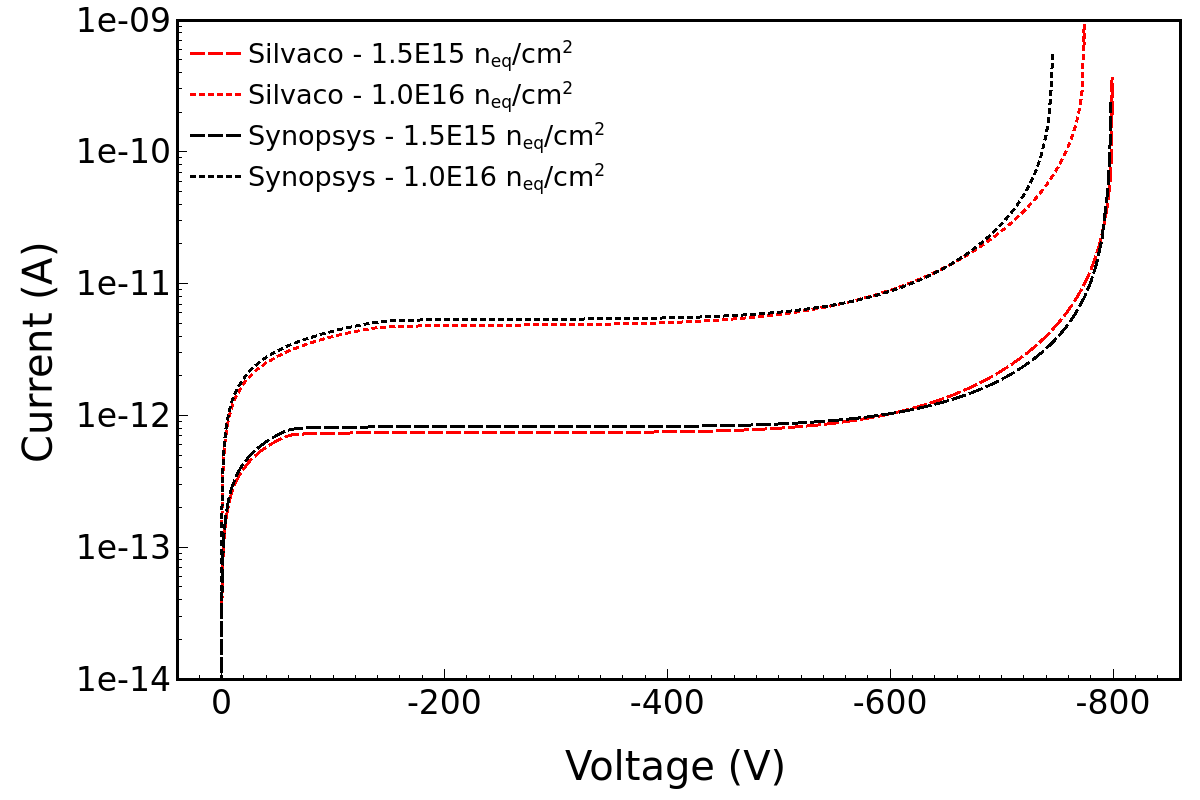}
   \includegraphics[width=0.49\textwidth]{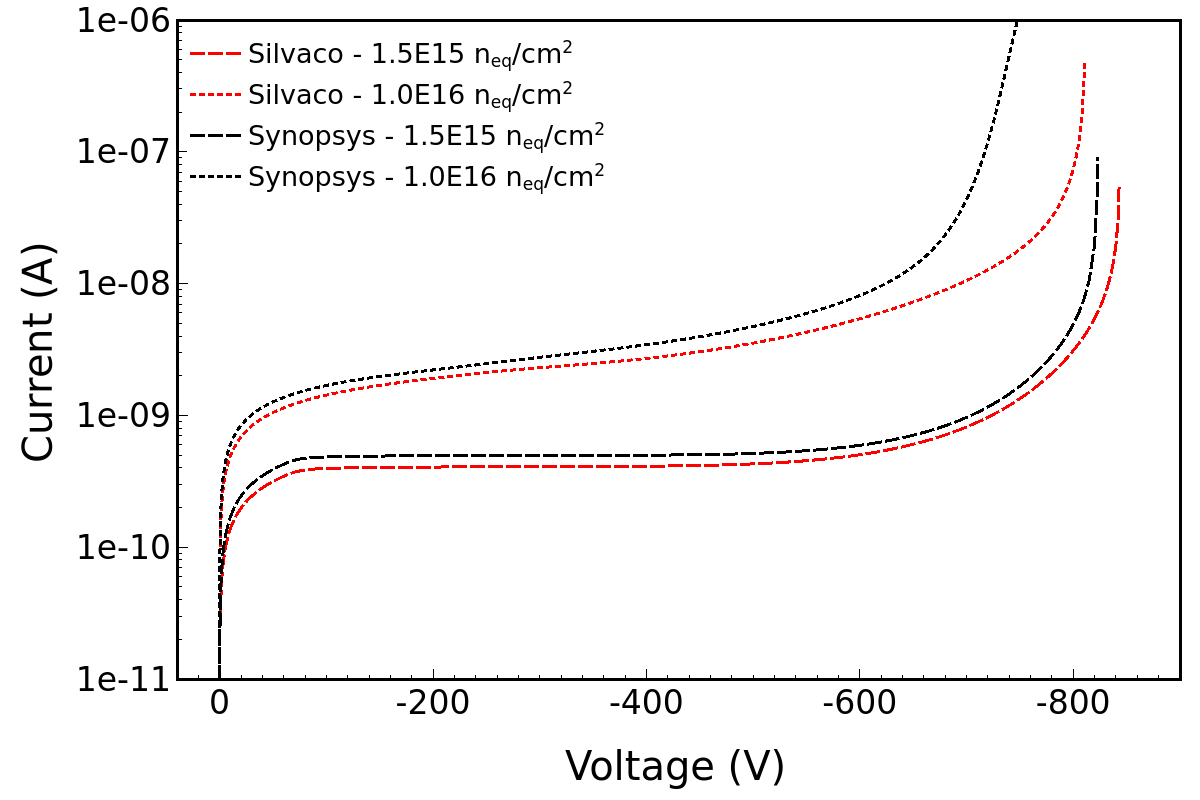} % requires the graphicx package
   \caption{\label{fig:irrIV}IV characteristics, results from irradiated device. (left) T~=~248~K, (right) T~=~300K.}
\end{figure}

In Figure~\ref{fig:irrIV} the IV characteristics of the irradiated diode are presented for all scenarios. 

At $T = $~248~K the agreement between the two TCAD tools is remarkably good; only at very large voltages the two predictions depart one from the other, especially at the largest fluence. This could be 
linked to the same issue observed before irradiation, {\it i.e.} being due to the different mesh grids between the two tools. 
At the highest temperature the agreement is more at a qualitative level, in particular at the largest fluence. A different impact ionization model was tested (by Okuto and Crowell~\cite{OKUTO1975161}) and 
 the level of disagreement was found to be even larger. This result could be investigated more but at the same time it is not so worrying due to the 
limited applicability of the results since a device after being irradiated at  $\Phi = 1.0\times10^{16}$~\SI{}{n_{eq}cm^{-2}} is very rarely operated at temperature above~0~C. For this reason in the following 
only results at $T = $~248~K will be presented.

\begin{figure}[htbp]
   \centering
   \includegraphics[width=0.49\textwidth]{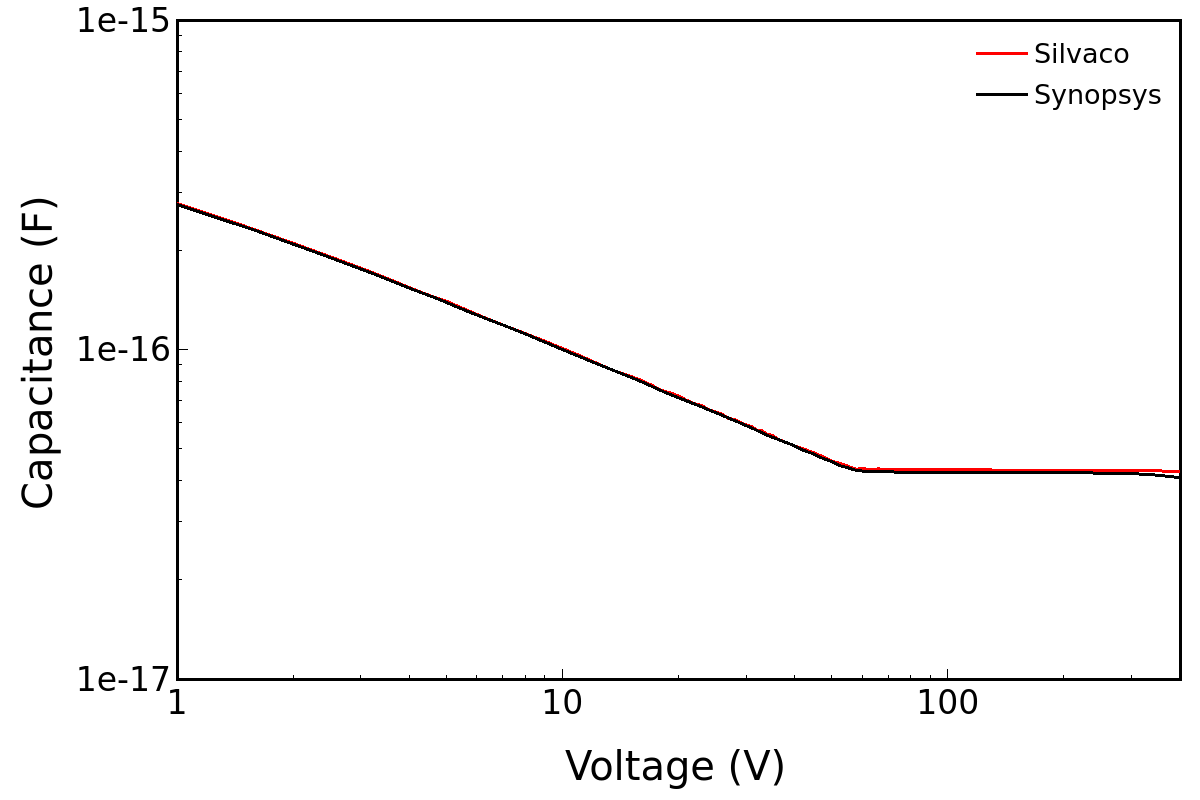}
   \includegraphics[width=0.49\textwidth]{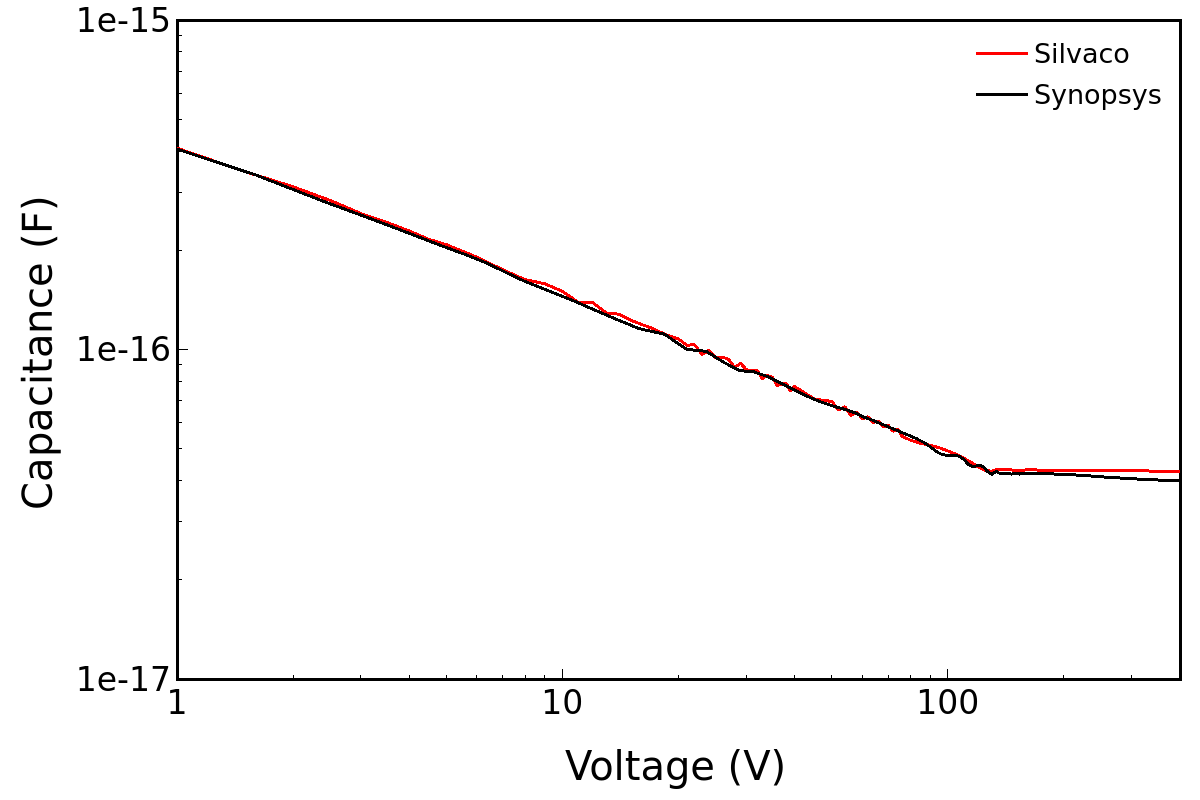} % requires the graphicx package
   \caption{\label{fig:irrCV}CV characteristics at $T = 248$~K, results from irradiated device. (left) $\Phi = 1.5\times10^{15}$~\SI{}{n_{eq}cm^{-2}}   (right) $\Phi = 1.0\times10^{16}$~\SI{}{n_{eq}cm^{-2}}.}
\end{figure}

In Figure~\ref{fig:irrCV} the CV characteristics at $T = 248$~K  of the irradiated device are reported when both bulk and interface defects are simulated. The level of agreement between the two tools is practically 
perfect.

\begin{figure}[htbp]
   \centering
   \includegraphics[width=0.32\textwidth]{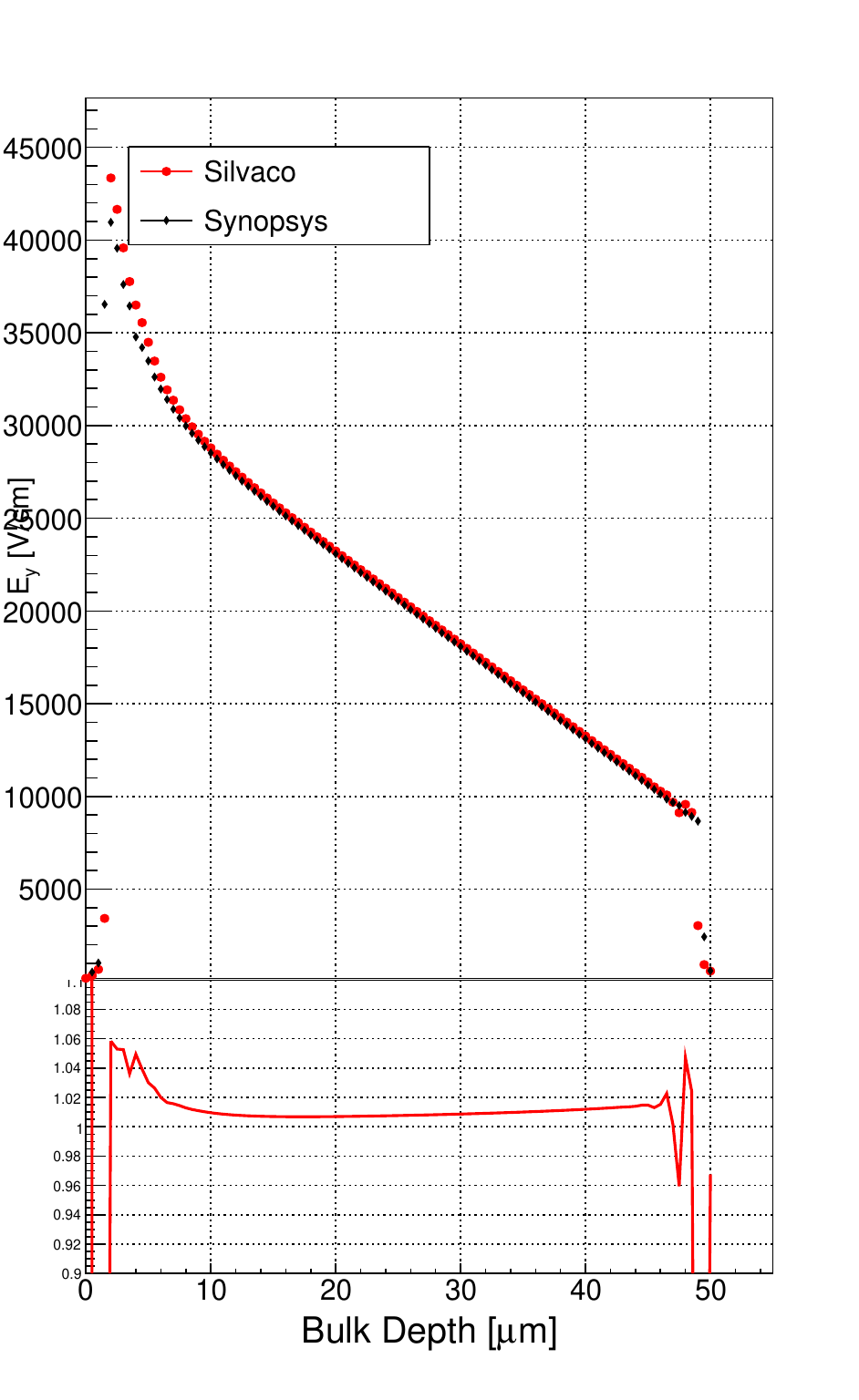}
   \includegraphics[width=0.32\textwidth]{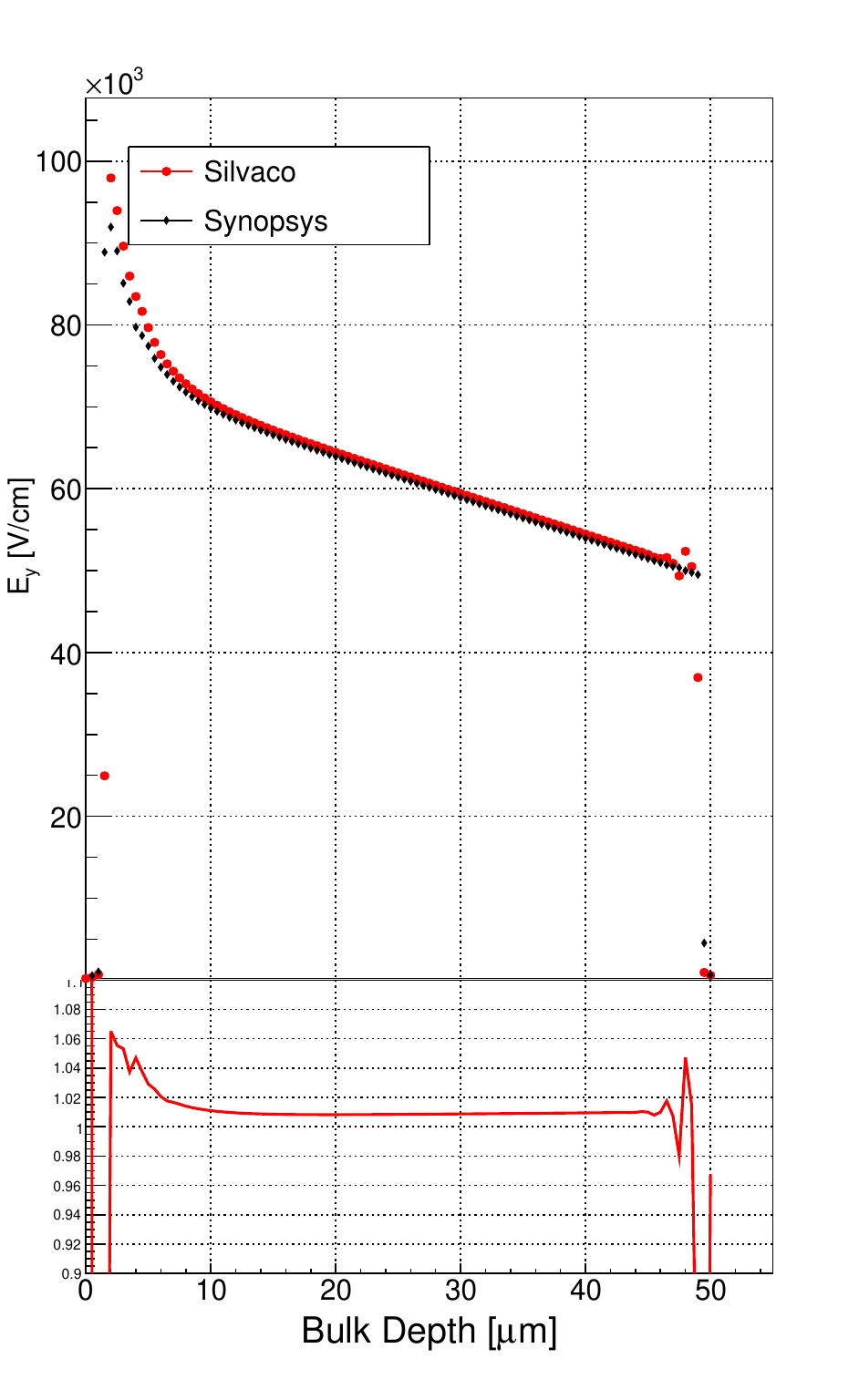} % requires the graphicx package
   \includegraphics[width=0.32\textwidth]{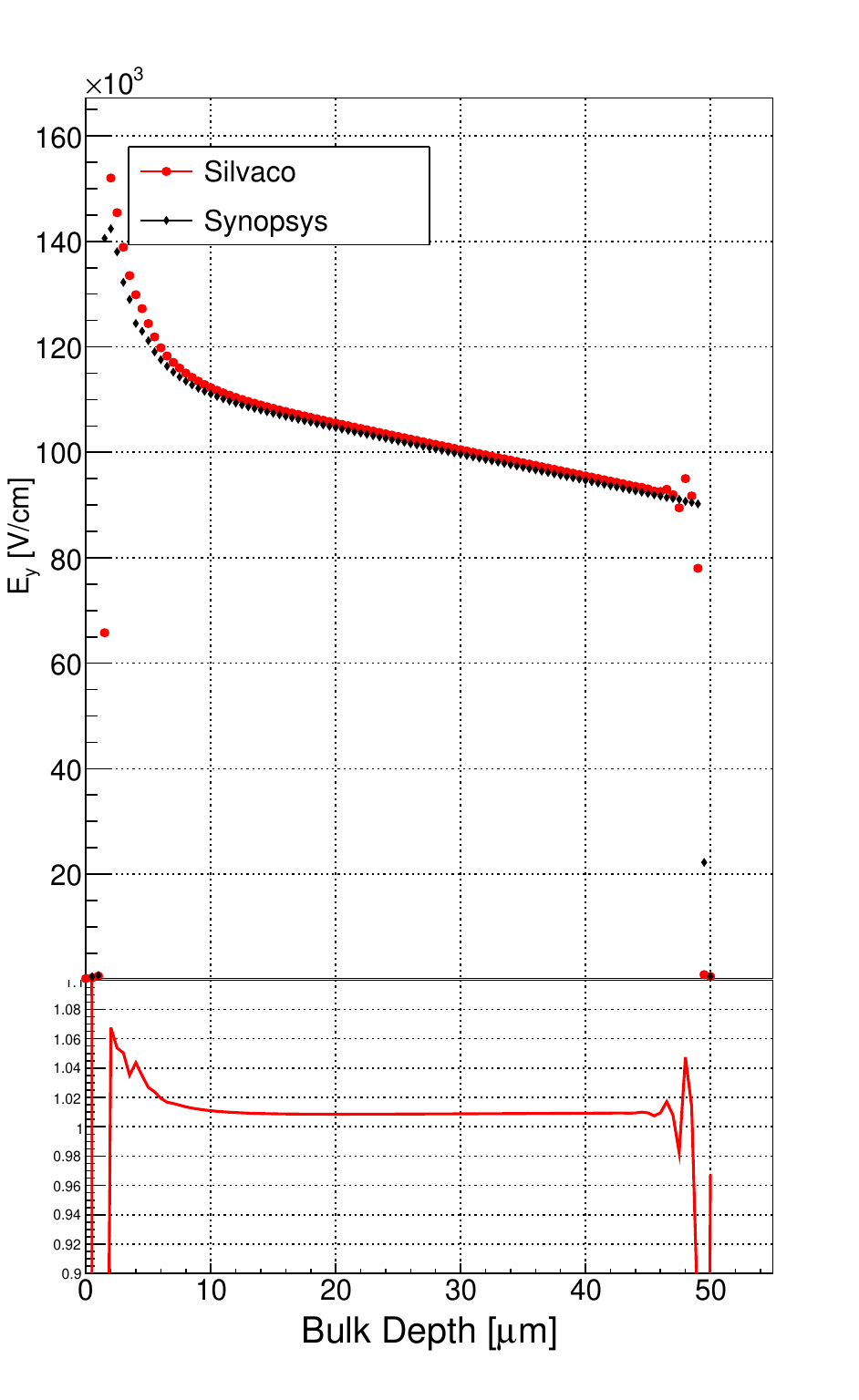} % requires the graphicx package
   \caption{\label{fig:irrEfieldLow}Electric field profile vs bulk depth  at $T = 248$~K, results from irradiated device at  $\Phi = 1.5\times10^{15}$~\SI{}{n_{eq}cm^{-2}} at different bias voltages: (left) 100~V, (mid) 300~V, 
   (right) 500~V.}
\end{figure}

In Figure~\ref{fig:irrEfieldLow} the simulated electric field profile as a function of the bulk depth is presented for a device irradiated at $\Phi = 1.5\times10^{15}$~\SI{}{n_{eq}cm^{-2}} for three different 
bias voltage values.

\begin{figure}[htbp]
   \centering
   \includegraphics[width=0.32\textwidth]{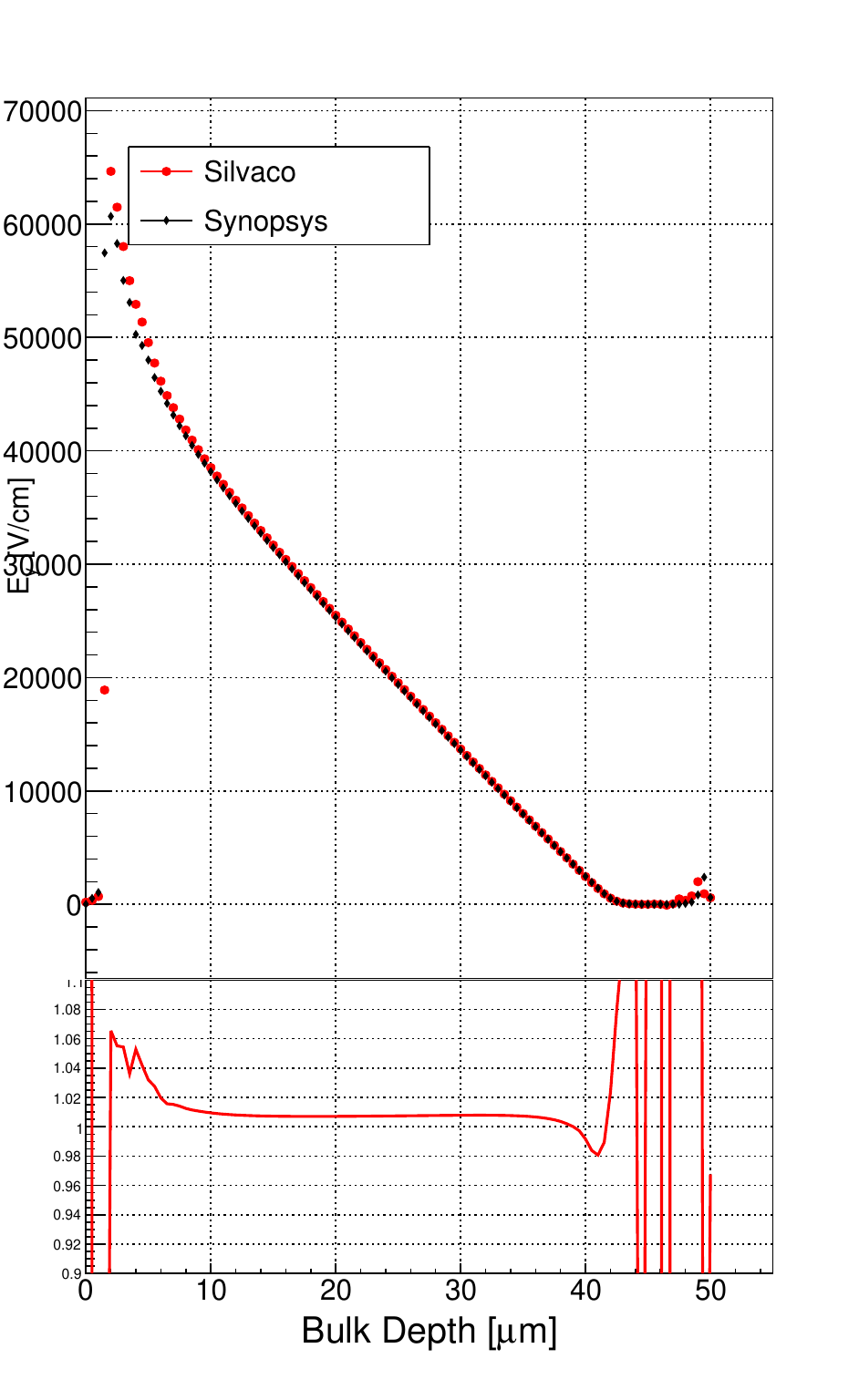}
   \includegraphics[width=0.32\textwidth]{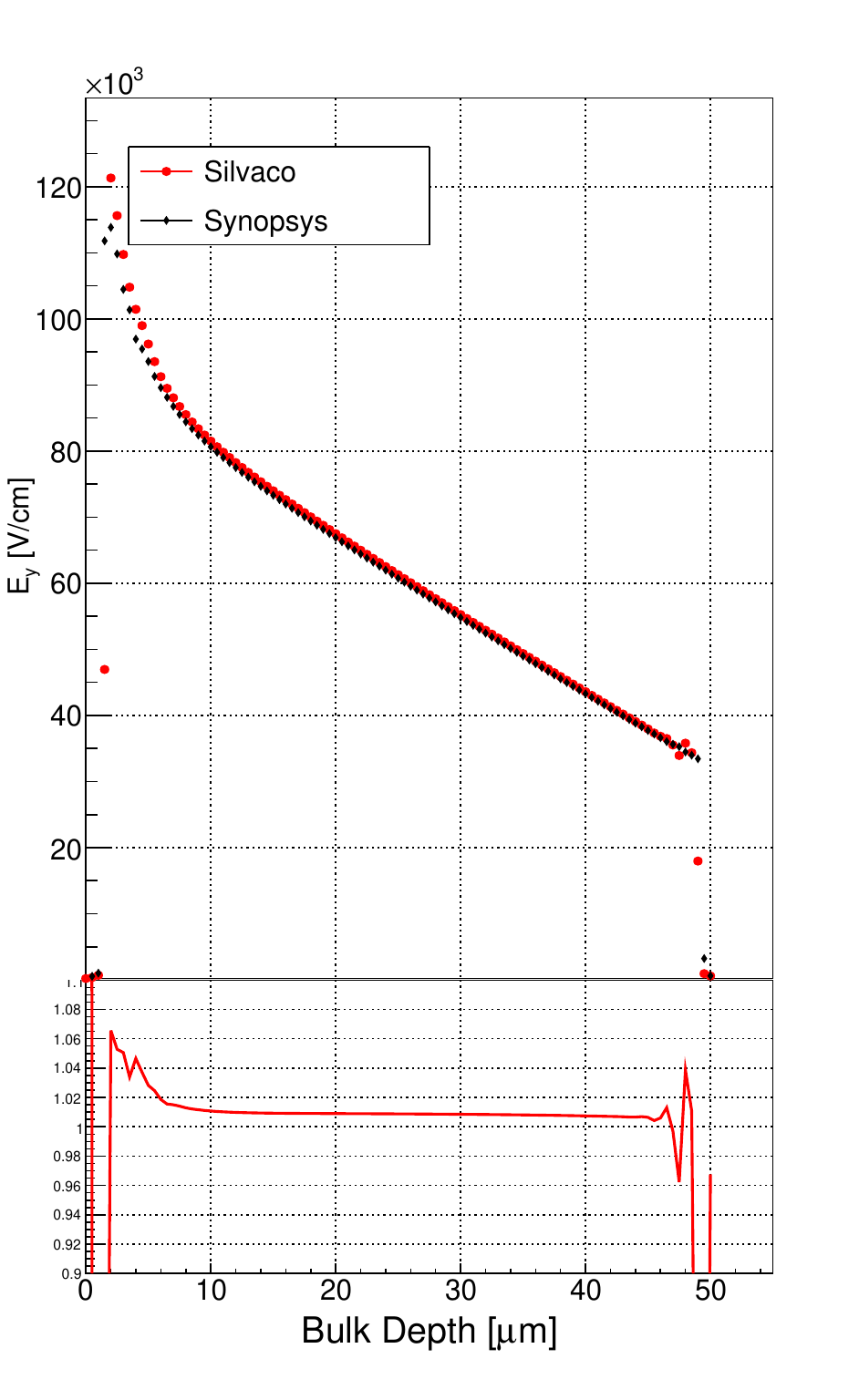} % requires the graphicx package
   \includegraphics[width=0.32\textwidth]{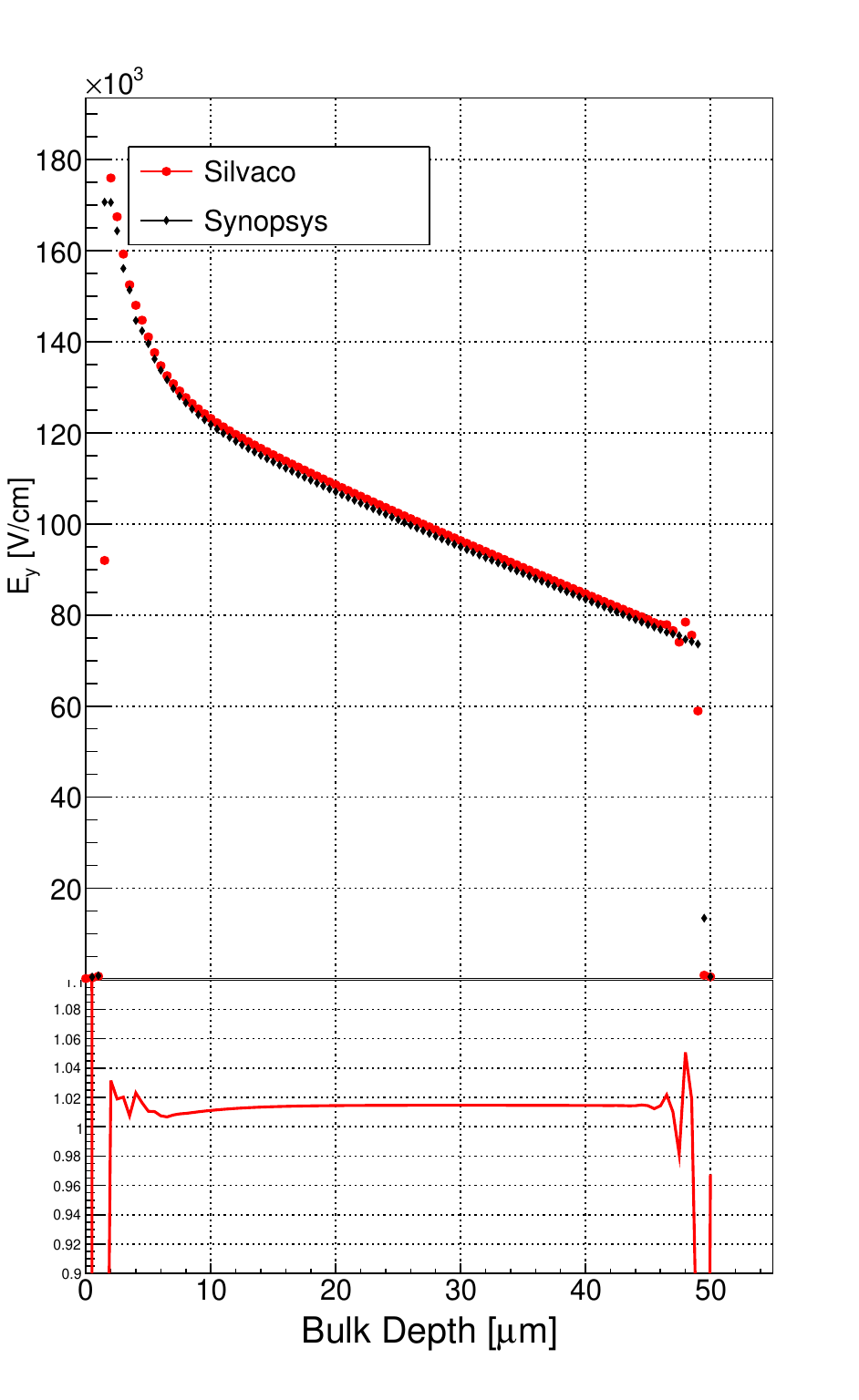} % requires the graphicx package
   \caption{\label{fig:irrEfieldHigh}Electric field profile vs bulk depth  at $T = 248$~K, results from irradiated device at  $\Phi = 1.0\times10^{16}$~\SI{}{n_{eq}cm^{-2}} at different bias voltages: (left) 100~V, (mid) 300~V, 
   (right) 500~V.}
\end{figure}

In Figure~\ref{fig:irrEfieldHigh} the simulated electric field profile as a function of the bulk depth is presented for a device irradiated at $\Phi = 1.0\times10^{16}$~\SI{}{n_{eq}cm^{-2}} for three different 
bias voltage values. Ratio of Silvaco over Synopsys results are presented at the bottom of each plot.

For both fluences and at all voltages inspected the level of agreement between the two tools is excellent in the bulk region, down to 1\%. The regions of front and backside implants show a lesser level of 
agreement, again probably due to mismatch in meshes between the tools and related differences in implants modelling.

\begin{figure*}[t] % Use figure* if you are in a two-column document
     \centering
     % --- Plot A ---
     \begin{subfigure}[b]{0.24\textwidth}
         \centering
         \includegraphics[width=\textwidth]{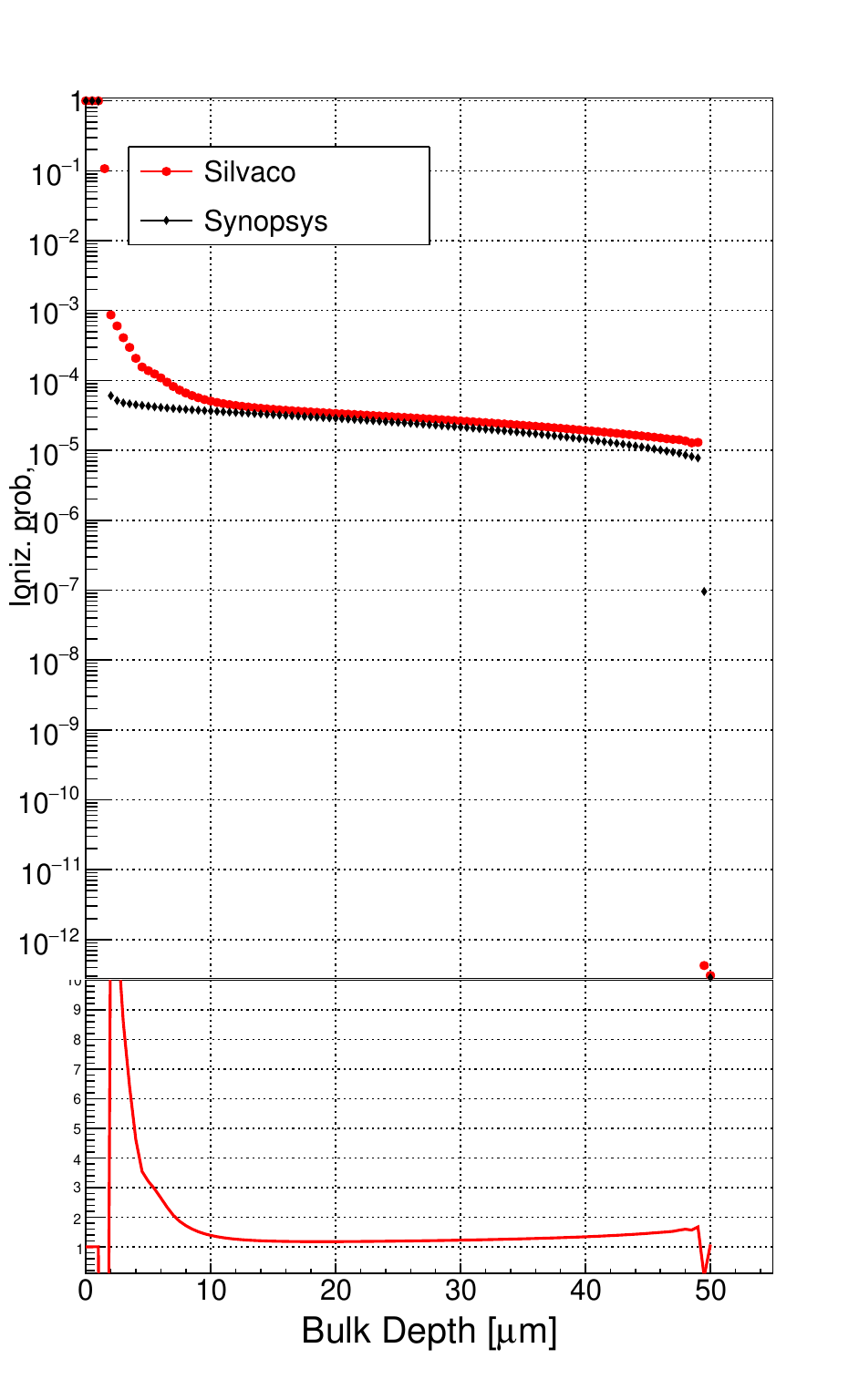}
         \caption{}
         \label{fig:plot_a}
     \end{subfigure}
     \hfill
     % --- Plot B ---
     \begin{subfigure}[b]{0.24\textwidth}
         \centering
         \includegraphics[width=\textwidth]{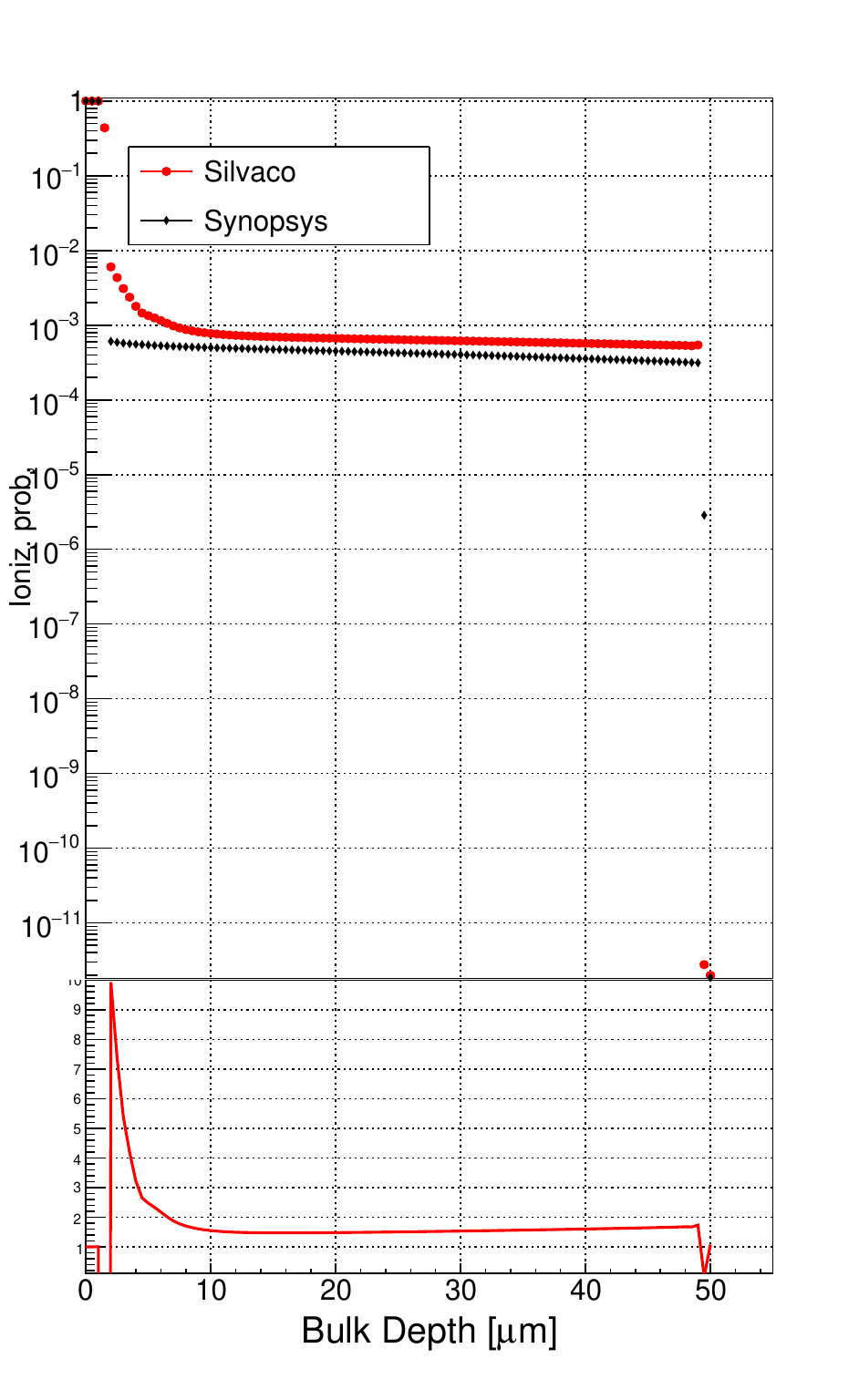}
         \caption{}
         \label{fig:plot_b}
     \end{subfigure}
     \hfill
     % --- Plot C ---
     \begin{subfigure}[b]{0.24\textwidth}
         \centering
         \includegraphics[width=\textwidth]{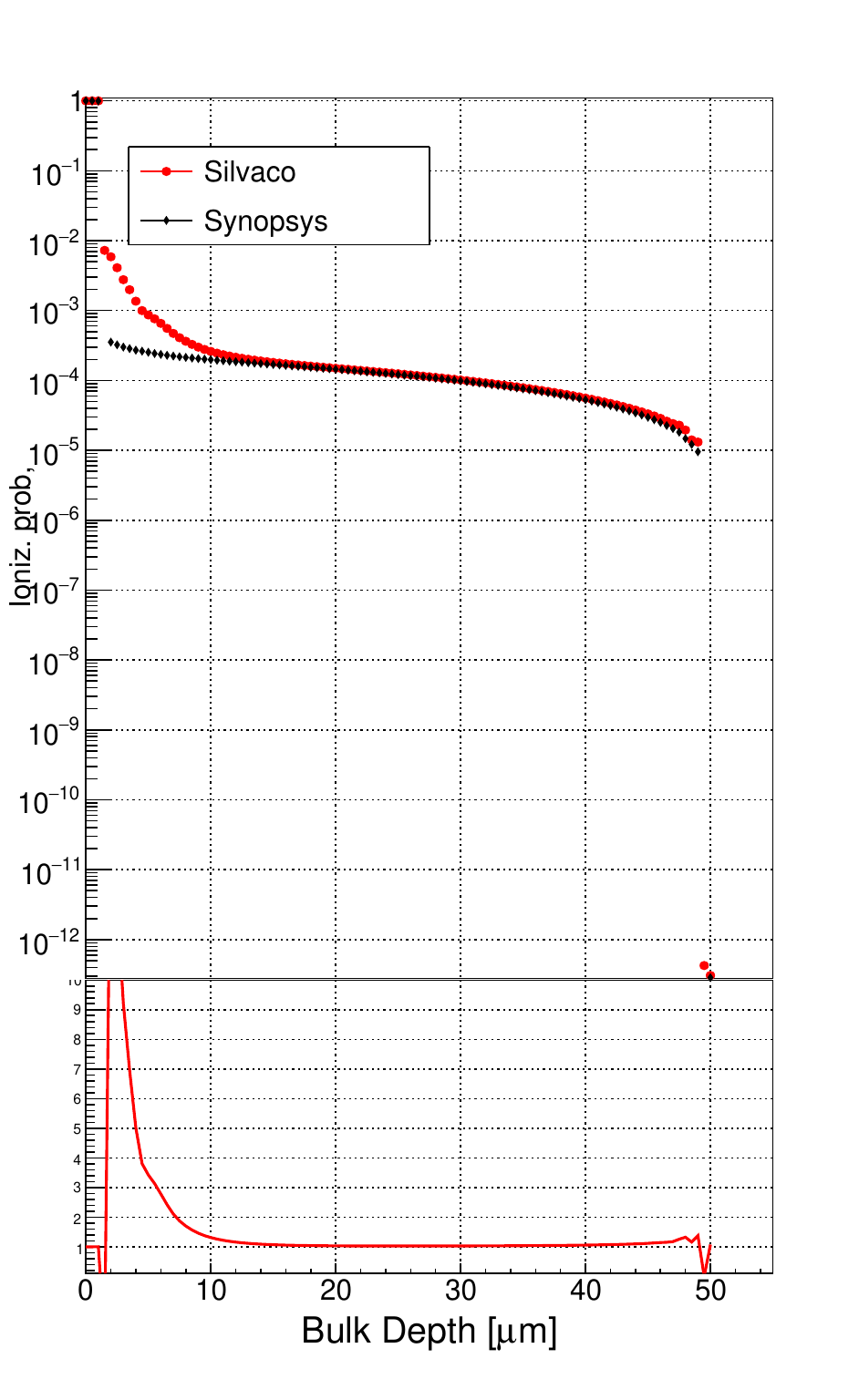}
         \caption{}
         \label{fig:plot_c}
     \end{subfigure}
     \hfill
     % --- Plot D ---
     \begin{subfigure}[b]{0.24\textwidth}
         \centering
         \includegraphics[width=\textwidth]{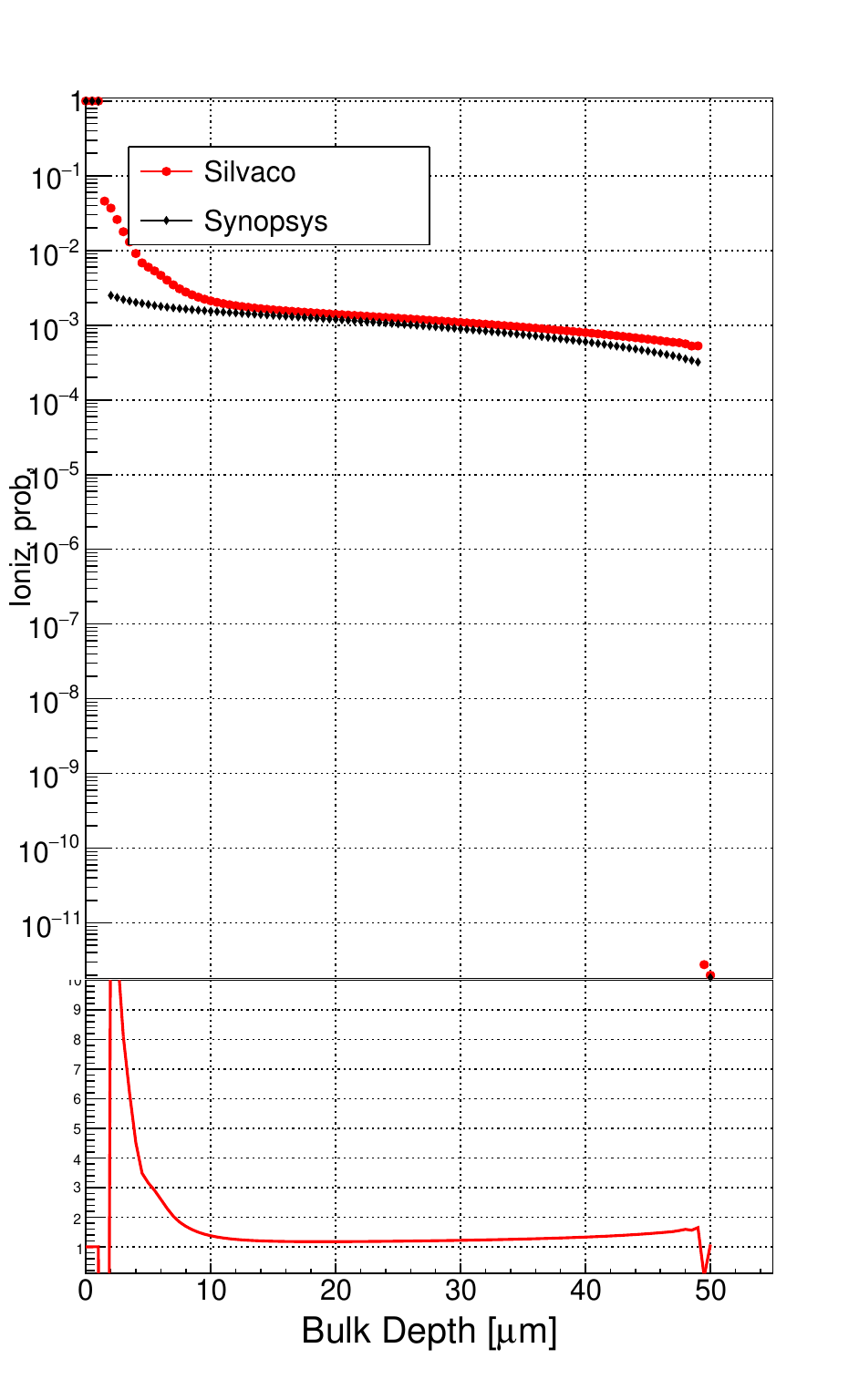}
         \caption{}
         \label{fig:plot_d}
     \end{subfigure}
         \caption{\label{fig:irrTraps}Trap ionisation probabilities of bulk defects at $T = 248$~K, results from irradiated device at 500~V. (a) and (b) $\Phi = 1.5\times10^{15}$~\SI{}{n_{eq}cm^{-2}}; 
         (c) and (d) $\Phi = 1.0\times10^{16}$~\SI{}{n_{eq}cm^{-2}}, (a) and (c) acceptor trap 1;  (b) and (d) acceptor trap 2.}
     
\end{figure*}

Given the overall excellent agreement between the two tools when simulating radiation effects it is interesting to investigate the ionisation probabilities of bulk defects.
In Figure~\ref{fig:irrTraps} the ionisation probabilities of bulk defects as a function of the bulk depth are presented at $T = $~248~K for both fluences. The results are reported for the acceptor trap 1 and 2
(0.42 and 0.46~eV below the lower edge of the conduction band) at 500~V.

The two tools agree within an order of magnitude in ionisation probabilities, down to 20\% in the bulk region. This level of agreement is indeed satisfactory given that electric field profiles agree at percent level.

\section{Conclusions and Outlook}
\label{sec:concl}

In this paper the compatibility of Silvaco and Synopsys TCAD tools results have been reported when simulating irradiated diodes at fluences relevant for the high-luminosity phase of LHC. The level of 
agreement is very good for crucial observables like leakage current and electric field. Slightly suboptimal agreement in breakdown voltage, electric field and trap ionisation probabilities 
close to implants could be due to the different mesh used with the different TCAD softwares. In the future similar studies on collected charge will be carried out. 
This study affirms the validity of the University of Perugia TCAD radiation damage model even on a different TCAD platform.
%\appendix
%\section{Some title}
%Please always give a title also for appendices.

\acknowledgments
The author(s) would like to acknowledge the use of the Gemini AI (Google) for assistance with language refinement and technical formatting of the manuscript.

\bibliographystyle{JHEP}
 \bibliography{mybiblio.bib}


%\begin{thebibliography}{99}

%\bibitem{a}
%Author, \emph{Title}, \emph{J. Abbrev.} {\bf vol} (year) pg.

%\bibitem{b}
%Author, \emph{Title},
%arxiv:1234.5678.

%\bibitem{c}
%Author, \emph{Title},
%Publisher (year).

% Please avoid comments such as "For a review'', "For some examples",
% "and references therein" or move them in the text. In general,
% please leave only references in the bibliography and move all
% accessory text in footnotes.

% Also, please have only one work for each \bibitem.

%\end{thebibliography}
\end{document}